\documentclass[sts]{imsart}

\RequirePackage{amsthm,amsmath,amsfonts,amssymb}
\RequirePackage[numbers,sort&compress]{natbib}
\RequirePackage[colorlinks,citecolor=blue,urlcolor=blue]{hyperref}
\RequirePackage{graphicx}

\startlocaldefs
\theoremstyle{plain}

\theoremstyle{remark}

\endlocaldefs

\begin{document}

\begin{frontmatter}
\title{Protocols for Observational Studies: An Application to Regression Discontinuity Designs}
%\title{A Sample Article Title with Some Additional Note\thanksref{t1}}
\runtitle{Protocols for Observational Studies: RD Designs}
%\thankstext{T1}{A sample additional note to the title.}

\begin{aug}
\author[A]{\fnms{Matias D.}~\snm{Cattaneo}\ead[label=e1]{cattaneo@princeton.edu}\orcid{0000-0003-0493-7506}}
\and
\author[B]{\fnms{Roc\'{i}o}~\snm{Titiunik}\ead[label=e2]{titiunik@princeton.edu}\orcid{0000-0001-5145-3059}}

\address[A]{Matias D. Cattaneo is Professor,
            Department of Operations Research and Financial Engineering,
            Princeton University,
            New Jersey, USA\printead[presep={\ }]{e1}.\newline}

\address[B]{Roc\'{i}o Titiunik is Professor,
            Department of Politics,
            Princeton University,
            New Jersey, USA\printead[presep={\ }]{e2}.}

\end{aug}

\begin{abstract}
    In his 2022 IMS Medallion Lecture delivered at the Joint Statistical Meetings, Prof. Dylan S. Small eloquently advocated for the use of protocols in observational studies. We discuss his proposal and, inspired by his ideas, we develop a protocol for the regression discontinuity design.
\end{abstract}


\end{frontmatter}

\section{Introduction}

A protocol, sometimes referred to as a ``pre-registration plan'', is a detailed outline of a research study that specifies hypotheses, outcomes, and analyses in advance. The use of protocols in randomized controlled trials (RCTs) is common and even required in certain circumstances. However, in observational studies, which are studies where the treatment of interest is neither randomized nor under the control of the researcher, the use of protocols is rare and undisciplined. Because the intervention occurs before the study is conceived and conducted, it is impossible to pre-specify a plan of analysis before the outcomes are realized and the data is available. At first glance, the approach to reserve protocols for RCTs seems reasonable, as RCTs are the only context where protocols can be designed before the intervention to offer a credible commitment device for empirical work.

In his insightful lecture, Prof. Dylan S. Small \cite{Small2024-STS} challenges this notion, rejecting the premise that protocols are only useful before the intervention when the opportunities for cheating are minimal. In his own words, ``the goal of an observational study protocol is not to protect against dishonest investigators but to aid honest investigators to do good science'' (p. 10). His proposed approach to the use of protocols has the potential to improve both the credibility and the effectiveness of a plethora of scientific findings, well beyond the specialized case of RCTs.

Because in an observational study the treatment assignment mechanism is unknown, researchers must invoke an assumption that allows for the identification of causal effects. One of the most common is the ignorability assumption, which postulates that the treatment is independent of the potential outcomes after conditioning on a set of observed confounders or covariates, together with a common support condition. Prof. Small proposes a protocol for the analysis of observational studies based on ignorability. His proposal calls for the transparent specification of the following items: (i) the study population, including which subjects will be included and excluded; (ii) the treatment, including which subjects will be included in the treated and control groups; (iii) the primary and secondary outcomes, including specific time periods for measurement in case of follow-up outcomes; (iv) the covariates that will be adjusted for; (v) the statistical methods for adjustment and analysis; and (vi) and the falsification and sensitivity analyses. 

Through a series of examples, Prof. Small discusses these elements and outlines a set of best practices to deal with various issues, including multiplicity of outcomes and subgroup analyses. He also considers the timing of the protocol, discussing whether it should be written before or after matching on the covariates.

Inspired by his passionate defense of observational study protocols and his examples on how to implement such protocols when employing methods based on ignorability, we demonstrate the broad applicability of Prof. Small's ideas by developing a companion protocol for the regression discontinuity (RD) design \citep{Cattaneo-Titiunik_2022_ARE}. Our discussion highlights both the similarities and differences between observational protocols for selection-on-observables and RD designs.

\section{A Protocol for RD Designs}

In the RD design, all units receive a score, and the treatment of interest is assigned to those units whose score values are above a certain cutoff, and not assigned to those units whose score values are below the cutoff. For example, municipalities may receive federal assistance if their poverty index is above a cutoff, and not receive any assistance otherwise. Because the distribution of the score is generally not known, the RD design is considered an observational study \citep{Sekhon-Titiunik_2016_ObsStud, Titiunik_2021_HandbookCh}, where the study of treatment effects thus depends on an external identifying assumption ensuring the ``comparability'' between treated and control groups. In the RD context, this assumption is that the subset of units assigned to treatment whose scores are close to the cutoff are comparable to the subset of units assigned to control whose scores are close to the cutoff. This key RD identifying assumption is often justified by the inability of units to systematically and precisely affect the value of the score that they receive, and is formalized either as continuity of the regression functions at the cutoff \citep{Hahn-Todd-vanderKlaauw_2001_ECMA} or as local randomization of the treatment assignment in a neighborhood of the cutoff \citep{Cattaneo-Frandsen-Titiunik_2015_JCI}. See \cite{Cattaneo-Titiunik-VazquezBare_2017_JPAM} for a comparison of these two RD frameworks.

\subsection{Basic elements of research design}

The foundation of a protocol for an RD design is the specification of its three main elements, and the rule that links them: the score that all units receive, the cutoff value, the treatment, and the specific rule that determines treatment assignment for units with scores above (or below) the cutoff. A key feature is that the RD design can only be invoked if an RD treatment assignment rule actually occurred. In this sense, the RD design exists independently of the researcher, and can be verified externally. This contrasts with other observational studies where external verifiability is not possible. For example, a study based on ignorability can compare smokers and non-smokers conditional on a set of covariates, but there is no external condition about the assignment mechanism that must be true in order to invoke the ignorability assumption. In general, researchers must postulate untestable causal mechanisms in order to justify the use of selection-on-observable methods. In contrast, an RD design cannot be assumed, it must occur first. This important feature of the RD design provides the foundation for many of the protocol items that follow.

\subsection{Study population}

Once the score, cutoff, and treatment assignment rule have been specified, the study population is easily identified by including all the units who received a score. This simplifies the criteria for defining the study population relative to studies based on ignorability. For example, in the Mountaintop Mining study described by Prof. Small, the treatment group is the Central Appalachian counties with a high amount of surface mining, and the control group is the other counties in the four states in which surface mining occurs. The union of the chosen treated and control groups constitutes the study population. But the choice of the control group might have been different, for example, if the researchers had decided to include counties in geographically similar neighboring states where surface mining does not take place. 

This ambiguity in the study population is minimized in an RD design, because all units who receive a score are automatically included in the study population, and all units without a score are excluded. At the same time, the RD design introduces a different type of ambiguity, because the study population is used to estimate a local parameter that only captures the causal effect of the treatment for units whose scores are close to the cutoff, and not for the entire population of units who receive scores. See \cite{Cattaneo-Keele-Titiunik_2023_SIM} for more discussion.

\subsection{Treatment}

Because RD designs often arise due to the implementation of social interventions according to a specific assignment rule, the treatment of interest is unambiguously specified: it is the intervention that is given to units with scores above (or below) the cutoff. In other words, in most RD designs, the intervention is clearly specified as part of the assignment rule, and is often explicitly written in programs, rules, or laws. This clarity does not always occur in other types of observational studies. Continuing with the Mountaintop Mining example, the treatment of interest is ``high and sustained'' surface mining, which requires the researchers to specify the levels of mining that will be considered high and sustained. Such ambiguity is typically removed from an RD design protocol, because the RD rule will explicitly say what counts as treatment.

Although the RD design clearly determines what the intervention is, this does not mean that this intervention is scientifically relevant. A well specified protocol will carefully explain what features of the intervention assigned with an RD rule are related to open scientific questions, and how the intervention is related to the treatment of scientific interest. This will require a specification of the causal path and the particular mechanisms by which the RD intervention is related to a treatment of interest, and how this treatment is related to outcomes of interest. In other words, the occurrence of an RD design does not, by itself, justifies its use as a scientific study. An RD protocol should include a justification of the scientific value of studying the proposed RD design.

\subsection{Outcomes}

In all experimental or non-experimental settings, researchers must have a prior scientific model linking the treatment to certain outcomes based on hypothesized causal mechanisms. This causal model bears no relation to the RD treatment assignment rule. As in other observational studies, the choice of primary and secondary outcomes in an RD design will depend on the scientific theories and hypotheses guiding the research program. On this point, observational studies based on ignorability and RD designs are similar, and the ideas of Prof. Small apply equally.

\subsection{Statistical Methods}

For the statistical analysis, researchers should start by specifying whether they will adopt a continuity-based or a local randomization analysis, or both. See \cite{Cattaneo-Idrobo-Titiunik_2020_Vol1, Cattaneo-Idrobo-Titiunik_2024_Vol2} for a practical introduction to the two approaches. If they adopt both, they should indicate which one of the two analyses will be the primary analysis, and which one will be secondary. 

We distinguish two prototypical cases. The first case arises when (i) the running variable is continuous, that is, the sample contains no (or very few) units with the same value of the score, and (ii) there are enough observations on either side of the cutoff. In this case, the primary RD analysis is based on continuity because the local randomization approach imposes stronger assumptions. The protocol should specify the continuity-based statistical methods that will be used to fit the regression functions and to perform inferences. A key item is how the regression functions will be estimated; in particular, for the case of local polynomial methods, the protocol should specify how the bandwidth will be selected, the order of the polynomial, the kernel function, how misspecfication error will be handled to obtain valid inferences, and how uncertainty will be quantified. The protocol should also specify whether a complementary local randomization analysis will be included to asses robustness of the results. 

The second case arises when continuity-based methods are not directly applicable in the absence of additional assumptions. This case includes two types of designs: (i) RD designs where the running variable is discrete, that is, the sample contains many observations that share the same value of the score, and there are few unique score values; and (ii) RD designs where the score is continuous but there are too few observations to use continuity-based methods. In this case, the primary analysis is naturally based on a local randomization approach. The protocol should specify how the window around the cutoff will be selected, what test statistics will be used to test the hypothesis of no effect, and whether inferences will be based on Fisherian (randomization inference methods) or large-sample approximations (Neyman or other infinite population methods). The protocol should also specify the hypothesized assignment mechanism within the window, including whether regression adjustments will be used.

\subsection{Pre-intervention Covariates}

A pre-treatment covariate is a characteristic that is measured before the intervention is assigned. In observational studies based on ignorability, covariates are essential for identification of the parameter of interest. Although treated and control units are often systematically different, ignorability postulates that they become comparable once we condition on covariates. Identifying the set of covariates that are needed for identification is one of the most important aspects of a protocol based on ignorability, and should be based on the underlying scientific (causal) model guiding the research question.

In contrast, in the RD design, identification of the effects of interest relies on the assumption that treated units near the cutoff are comparable to control units near the cutoff. Whether this assumption is formalized as continuity or as local randomization, the only covariate required for identification is the RD score, the main pre-intervention covariate in the RD design. Importantly, this formalization does not involve conditioning or matching on the score; in fact, because of the lack of common support in the score for units above versus below the cutoff, it is impossible to use ignorability in RD designs in the absence of additional, more restrictive assumptions.

However, it is common for researchers to have available other pre-intervention covariates in addition to the RD score. These additional covariates have three potential roles: falsification checks, efficiency gains, and heterogeneity analyses. The most important use of additional covariates in the RD design is for falsification purposes. If it is indeed true that units barely assigned to treatment are comparable to units barely assigned to control, it should be the case that these two groups are similar in terms of pre-determined covariates. To asses this empirically, all RD protocols should specify a list of covariates that will be analyzed as outcomes. The expectation is that the null hypothesis that the treatment effect is zero will fail to be rejected. A protocol should specify how many and which imbalanced covariates will be tolerated, and a procedure for what to do in case this benchmark is exceeded.

The second role for covariates in an RD protocol is related to efficiency. Pre-determined covariates can be included in the analysis to reduce the variability of the estimates and thus increase precision. For example, local polynomial methods can be augmented to include pre-determined covariates in addition to the score, and local randomization methods can be implemented with covariate adjustment. An RD protocol should specify whether covariate adjustment will be included in the analysis for efficiency purposes, how it will be implemented, and which covariates will be used in order to mitigate the perception of p-hacking. Importantly, the protocol should also specify how statistical disagreements between unadjusted and covariate-adjusted RD estimates will be resolved. 

Covariates can also be used to explore heterogeneity by studying RD treatment effects for different subgroups of the population or considering interaction effects. As in the case of protocols based on ignorability discussed by Prof. Small, the RD protocol should specify any heterogeneity analyses, clearly indicating which subgroups or interactions will be analyzed.

Finally, unlike the case of other observational studies, covariates cannot be used to ``fix'' an RD design where units barely above the cutoff are different from units barely below the cutoff. Unless researchers are willing to make additional assumptions, covariates cannot be used to ``control'' for systematic differences between treated and control groups, much like covariates cannot be used to fix a randomized controlled experiment where the treatment and control groups are systematically different due to chance imbalances. See \citep{Cattaneo-Keele-Titiunik_2023_HandbookCh} for more discussion.

\subsection{Falsification and Sensitivity}

Falsification analyses are a fundamental part of any observational study. Because the assumption that units assigned to treatment are comparable to units assigned to control is not true by construction, all protocols for observational studies should include a plan to conduct empirical analyses that assess, to the extent possible, the plausibility of the underlying identifying assumptions. 

Some falsification analyses are generic and apply to all observational studies, while others are specific to the RD design. The most important generic falsification analysis is an analysis of variables that, given the causal mechanism hypothesized, should not be affected by the intervention or treatment of interest. These variables include pre-determined covariates that capture important characteristics of the units, as well as variables that, despite being determined after the treatment, should not be affected by it---usually called negative control outcomes or placebo outcomes. An RD protocol should include the list of covariates and negative control outcomes that will be included, and how inferences will be conducted.

An RD protocol should also specify whether RD-specific falsification analyses will be included. These include density analyses that test whether the number of observations just above the cutoff is similar to the number of observations just below the cutoff, analyses that vary the choice of  bandwidth or window, and analyses that leave out the observations closest to the cutoff. These tests are specific to the RD design because they exploit the RD principle of focusing on units close to the cutoff. The protocol should provide details about how these tests will be implemented.

As a complement to falsification tests, sensitivity analyses methods investigate whether and how much the conclusions of a study would change if the main assumptions were violated. For example, Prof. Small discusses Rosenbaum's $\Gamma$ sensitivity method. This method is directly applicable in RD designs based on the local randomization approach. 

\subsection{Multiplicity Corrections}

In any protocol, the outcome of interest should be carefully chosen based on the scientific question and the causal mechanisms that are part of the theoretical justification for the design. If the underlying scientific theory suggests that many outcomes are equally important, all of these outcomes can be analyzed. In this case, inference corrections will be needed, and the protocol should specify the parameters of the correction. Otherwise, researchers should use their scientific theories as guidance to select the most important outcome, and designate it as the primary outcome in the protocol.

Prof. Small discusses three strategies to deal with multiple outcomes in observational studies based on ignorability: (i) choosing a small number of primary outcomes a priori (i.e., one or two), (ii) testing many outcomes using a Bonferroni-type correction, and (iii) splitting the observations into a planning sample---used to choose which outcome is most associated with the treatment---and an independent analysis sample---used to analyze the treatment effect on the outcome chosen in the planning sample. He argues for the use of sample splitting for outcomes with high design sensitivity, because this leads to power gains in the analysis sample despite the loss in sample size.

We consider these strategies in the context of an RD design. Although sample splitting has advantages, in many RD applications it will not be a feasible method to deal with outcome multiplicity. RD designs are based on the assumption that, near the cutoff, units assigned to treatment are comparable to units assigned to control. Focusing the analysis on units with scores near the cutoff is thus essential in all RD analyses. This ``localizing'' process necessarily reduces the size of the sample effectively used. In continuity-based analyses, it is not uncommon for researchers to use an effective sample containing only about $10\%$ to $20\%$ of the total number of observations. In local randomization analyses, this is even more dramatic, with analyses sometimes discarding more than $95\%$ of the observations. Except in RD designs involving big data, sample splitting may not be practically feasible.

A natural alternative is to focus on the few most relevant outcomes based on the scientific theory underlying the RD design, pre-specifying them in the protocol as primary outcomes. In cases where the scientific hypothesis calls for considering many outcomes, a multiple testing correction such as Bonferroni should be used. In this case, the RD protocol should include the list outcomes and the particular multiplicity correction method that will be used. 

\section{Timing of RD protocol}

In RCTs, protocols are written before the treatment is assigned. Prof. Small asserts that, in matched observational studies, ``matching takes the place of random assignment'', which suggests that protocols should be written before matching. However, he describes a tension that is typical of most observational studies: researchers want a design that not only minimizes confounding or self-selection, but also is as informative as possible. In matched observational studies, these goals are in tension because reducing confounding calls for matching in as many covariates as possible, but matching on a highly imbalanced covariate can lead to efficiency losses. Writing the protocol after matching has the advantage that covariate balance information can be used to decide which covariates to match on, but it has the disadvantage that it can lead researchers to develop post hoc justifications for why certain covariates should be excluded. Regardless of whether the protocol is written before or after matching, Prof. Small takes for granted that the protocol should always be written before analyzing the outcomes of interest.

Like in any other observational study, the most important principle regarding the timing of an RD protocol is that the protocol should be written before analyzing the outcome. A secondary question is whether the protocol should be written before or after studying the density of the running variable and selecting the neighborhood where the analysis will take place. The answer depends on whether the RD design is analyzed using the local randomization or the continuity-based framework. In a local randomization framework, the window around the cutoff is chosen based on balance on pre-determined covariates. Because this window can be chosen without analyzing the outcome, the situation is analogous to the matching case, where the local units in the window behave as the matched units in observational studies based on ignorability. Thus, when a local randomization  framework is employed, the timing issues discussed by Prof. Small apply directly, with the caveat that the sample size will be considerably smaller and some strategies such as sample splitting may be infeasible. 

In the continuity case, the question is more complicated. Because in a continuity-based approach it will be impossible to choose the bandwidth without analyzing the outcome, an RD protocol that outlines a continuity-based analysis cannot include the specific bandwidth that will be used. There are two alternative options. One is to specify the method that will be used to select the bandwidth, but not the bandwidth itself. The other, available only when the sample size is very large, is to employ sample splitting and use the planning sample to select the optimal bandwidth by learning about the curvature of the regression functions and other features of the underlying data generating process. Then, the separate analysis sample is used to implement the RD analysis with the bandwidth chosen in the planning sample. This approach would require an adjustment to reflect the potentially different sample sizes. For example, if the planning sample is of size $n_1$ and the analysis sample is of size $n_2$, the mean squared error (MSE) optimal bandwidth for local polynomial analysis in the planning sample will be $h_1^\mathtt{MSE} \propto n_1^{-1/(3+2p)}$, where $p$ is the polynomial order. The MSE-optimal bandwidth in the planning sample can be obtained as $h_2^\mathtt{MSE} = h_1^\mathtt{MSE} \cdot (n_1/n_2)^{1/(3+2p)}$. The bandwidth $h_2^\mathtt{MSE}$ thus uses the estimated curvature and variability from the planning sample, and then adjusts it to reflect the number of observations in the analysis sample. This procedure allows researchers to look at features of the outcome and possibly adjust the protocol in a more principled and objective way. See \cite{Calonico-Cattaneo-Farrell_2020_ECTJ} for more discussion on bandwidth selection in RD designs.

In sum, an RD protocol based on the continuity framework will typically be written before selecting the specific bandwidth used for the analysis unless the sample size is very large and allows for sample splitting, while an RD protocol based on local randomization can be written before or after selecting the window around the cutoff. In both cases, if the protocol is designed after studying the distribution of the RD score in the absence of sample splitting, the analysis should be focused on the score only, and not include the outcomes of interest that will later be used for the main study. Inspection of the RD score can involve histograms and estimated density plots to asses how the score is distributed near the cutoff, but should not include any inspection of how the outcome and the score are related.

\section{Conclusion}

Prof. Dylan S. Small \cite{Small2024-STS} has made an landmark contribution by advocating for the use of protocols in observational studies. Thousands of key scientific questions can only be studied with observational studies because randomization of treatments is unfeasible or unethical. And even when randomized controlled experiments are possible, observational studies are often instrumental in providing additional evidence regarding the generalizability of experimental effects. Given the vast amount of scientific findings that are based on observational studies, the adoption of protocols advocated by Prof. Small has enormous potential to change society for the better. We enthusiastically endorse his call for the use of protocols in observational studies. We hope our discussion has made clear that protocols are not only useful in matched observational studies, but also in RD designs and, by analogy, in all other types of observational studies.

There is an added benefit to the use of protocols in observational studies. In an era of high political polarization, the legitimacy of science has been under attack. Some people may distrust scientific findings for political or ideological reasons, and some scientists may have strong preferences about the kinds of results that they hope to find, as Prof. Small forcefully illustrates with several examples from his own research. The widespread adoption of observational study protocols would increase the credibility of scientific findings and help restore and protect the legitimacy of science, which is essential to translate scientific advancement into societal benefits.

\begin{funding}
    The first author was supported by the National Science Foundation through grants DMS-2304646 and SES-2241575. The second author was supported by was supported by the National Science Foundation through grant SES-2241575.
\end{funding}

\bibliographystyle{imsart-number} % Style BST file (imsart-number.bst or imsart-nameyear.bst)
\bibliography{references}       % Bibliography file (usually '*.bib')

\begin{thebibliography}{12}
% BibTex style file: imsart-number.bst, 2017-11-03
% Default style options (sort=1,type=number).
% Used options (sort=1,type=number).

\bibitem{Calonico-Cattaneo-Farrell_2020_ECTJ}
\begin{barticle}[author]
\bauthor{\bsnm{Calonico},~\bfnm{Sebastian}\binits{S.}},
  \bauthor{\bsnm{Cattaneo},~\bfnm{Matias~D.}\binits{M.~D.}} \AND
  \bauthor{\bsnm{Farrell},~\bfnm{Max~H.}\binits{M.~H.}}
(\byear{2020}).
\btitle{Optimal Bandwidth Choice for Robust Bias Corrected Inference in
  Regression Discontinuity Designs}.
\bjournal{Econometrics Journal}
\bvolume{23}
\bpages{192--210}.
\end{barticle}
\endbibitem

\bibitem{Cattaneo-Frandsen-Titiunik_2015_JCI}
\begin{barticle}[author]
\bauthor{\bsnm{Cattaneo},~\bfnm{Matias~D.}\binits{M.~D.}},
  \bauthor{\bsnm{Frandsen},~\bfnm{Brigham}\binits{B.}} \AND
  \bauthor{\bsnm{Titiunik},~\bfnm{Rocio}\binits{R.}}
(\byear{2015}).
\btitle{Randomization Inference in the Regression Discontinuity Design: An
  Application to Party Advantages in the U.S. Senate}.
\bjournal{Journal of Causal Inference}
\bvolume{3}
\bpages{1--24}.
\end{barticle}
\endbibitem

\bibitem{Cattaneo-Idrobo-Titiunik_2020_Vol1}
\begin{bbook}[author]
\bauthor{\bsnm{Cattaneo},~\bfnm{Matias~D.}\binits{M.~D.}},
  \bauthor{\bsnm{Idrobo},~\bfnm{Nicol\'{a}s}\binits{N.}} \AND
  \bauthor{\bsnm{Titiunik},~\bfnm{Roc\'{i}o}\binits{R.}}
(\byear{2020}).
\btitle{A Practical Introduction to Regression Discontinuity Designs:
  Foundations}.
\bpublisher{Cambridge Elements: Quantitative and Computational Methods for
  Social Science, Cambridge University Press}.
\end{bbook}
\endbibitem

\bibitem{Cattaneo-Idrobo-Titiunik_2024_Vol2}
\begin{bbook}[author]
\bauthor{\bsnm{Cattaneo},~\bfnm{Matias~D.}\binits{M.~D.}},
  \bauthor{\bsnm{Idrobo},~\bfnm{Nicol\'{a}s}\binits{N.}} \AND
  \bauthor{\bsnm{Titiunik},~\bfnm{Roc\'{i}o}\binits{R.}}
(\byear{2024}).
\btitle{A Practical Introduction to Regression Discontinuity Designs:
  Extensions}.
\bpublisher{Cambridge Elements: Quantitative and Computational Methods for
  Social Science, Cambridge University Press}.
\end{bbook}
\endbibitem

\bibitem{Cattaneo-Keele-Titiunik_2023_SIM}
\begin{barticle}[author]
\bauthor{\bsnm{Cattaneo},~\bfnm{Matias~D.}\binits{M.~D.}},
  \bauthor{\bsnm{Keele},~\bfnm{Luke}\binits{L.}} \AND
  \bauthor{\bsnm{Titiunik},~\bfnm{Rocio}\binits{R.}}
(\byear{2023}).
\btitle{A Guide to Regression Discontinuity Designs in Medical Applications}.
\bjournal{Statistics in Medicine}
\bvolume{42}
\bpages{4484--4513}.
\end{barticle}
\endbibitem

\bibitem{Cattaneo-Keele-Titiunik_2023_HandbookCh}
\begin{bincollection}[author]
\bauthor{\bsnm{Cattaneo},~\bfnm{Matias~D.}\binits{M.~D.}},
  \bauthor{\bsnm{Keele},~\bfnm{Luke}\binits{L.}} \AND
  \bauthor{\bsnm{Titiunik},~\bfnm{Rocio}\binits{R.}}
(\byear{2023}).
\btitle{Covariate Adjustment in Regression Discontinuity Designs}.
In \bbooktitle{Handbook of Matching and Weighting in Causal Inference}
(\beditor{\bfnm{D.~S.~Small}\binits{D.~S.~S.}~\bsnm{J.~R.~Zubizarreta}
  \bsuffix{E.~A.~Stuart}} \AND
  \beditor{\bfnm{P.~R.}\binits{P.~R.}~\bsnm{Rosenbaum}}, eds.)
\bchapter{8},
\bpages{153--168}.
\bpublisher{Chapman \& Hall}, \baddress{Boca Raton, FL}.
\end{bincollection}
\endbibitem

\bibitem{Cattaneo-Titiunik_2022_ARE}
\begin{barticle}[author]
\bauthor{\bsnm{Cattaneo},~\bfnm{Matias~D.}\binits{M.~D.}} \AND
  \bauthor{\bsnm{Titiunik},~\bfnm{Rocio}\binits{R.}}
(\byear{2022}).
\btitle{Regression Discontinuity Designs}.
\bjournal{Annual Review of Economics}
\bvolume{14}
\bpages{821--851}.
\end{barticle}
\endbibitem

\bibitem{Cattaneo-Titiunik-VazquezBare_2017_JPAM}
\begin{barticle}[author]
\bauthor{\bsnm{Cattaneo},~\bfnm{Matias~D.}\binits{M.~D.}},
  \bauthor{\bsnm{Titiunik},~\bfnm{Rocio}\binits{R.}} \AND
  \bauthor{\bsnm{Vazquez-Bare},~\bfnm{Gonzalo}\binits{G.}}
(\byear{2017}).
\btitle{Comparing Inference Approaches for RD Designs: A Reexamination of the
  Effect of Head Start on Child Mortality}.
\bjournal{Journal of Policy Analysis and Management}
\bvolume{36}
\bpages{643--681}.
\end{barticle}
\endbibitem

\bibitem{Hahn-Todd-vanderKlaauw_2001_ECMA}
\begin{barticle}[author]
\bauthor{\bsnm{Hahn},~\bfnm{Jinyong}\binits{J.}},
  \bauthor{\bsnm{Todd},~\bfnm{Petra}\binits{P.}} \AND
  \bauthor{\bparticle{van~der} \bsnm{Klaauw},~\bfnm{Wilbert}\binits{W.}}
(\byear{2001}).
\btitle{Identification and Estimation of Treatment Effects with a
  Regression-Discontinuity Design}.
\bjournal{Econometrica}
\bvolume{69}
\bpages{201--209}.
\end{barticle}
\endbibitem

\bibitem{Sekhon-Titiunik_2016_ObsStud}
\begin{barticle}[author]
\bauthor{\bsnm{Sekhon},~\bfnm{Jasjeet~S.}\binits{J.~S.}} \AND
  \bauthor{\bsnm{Titiunik},~\bfnm{Roc\'{i}o}\binits{R.}}
(\byear{2016}).
\btitle{Understanding Regression Discontinuity Designs as Observational
  Studies}.
\bjournal{Observational Studies}
\bvolume{2}
\bpages{174--182}.
\end{barticle}
\endbibitem

\bibitem{Small2024-STS}
\begin{barticle}[author]
\bauthor{\bsnm{Small},~\bfnm{Dylan~S.}\binits{D.~S.}}
(\byear{2024}).
\btitle{Protocols for Observational Studies: Methods and Open Problems}.
\bjournal{Statistical Science}.
\end{barticle}
\endbibitem

\bibitem{Titiunik_2021_HandbookCh}
\begin{bincollection}[author]
\bauthor{\bsnm{Titiunik},~\bfnm{Rocio}\binits{R.}}
(\byear{2021}).
\btitle{Natural Experiments}.
In \bbooktitle{Advances in Experimental Political Science}
(\beditor{\bfnm{J.~N.}\binits{J.~N.}~\bsnm{Druckman}} \AND
  \beditor{\bfnm{D.~P.}\binits{D.~P.}~\bsnm{Green}}, eds.)
\bchapter{6},
\bpages{103-129}.
\bpublisher{Cambridge University Press}.
\end{bincollection}
\endbibitem

\end{thebibliography}

\end{document}